\documentstyle[12pt]{article}
\newcommand{\bm}[1]{\mbox{\boldmath $#1$}}
\newcommand{\lb}[1]{\label{#1}}
\newcommand{\bb}[1]{\bibitem{#1}}
\newcommand{\be}{\begin{equation}}
\newcommand{\ee}{\end{equation}}
\def\ba{\begin{eqnarray}}
\def\ea{\end{eqnarray}}
\def\ds{\displaystyle}

\def\n{\noindent}
\def\s{\sigma}
\def\k{\kappa}
\def\L{\Lambda}
\def\e{{\mathrm e}}

\begin{document}
\begin{center}

{\Large\bf The cosmological gravitating $\s$ model:} \\
{\Large\bf solitons and black holes} \\

\vskip1cm

{\bf G.\ Cl\'ement$^{(a)}$\footnote{e-mail: gclement@lapp.in2p3.fr}
and A. Fabbri$^{(b)}$\footnote{e-mail: fabbria@bo.infn.it}}

\end{center}

\medskip

{\small \n $^{(a)}$Laboratoire de Physique Th\'eorique LAPTH (CNRS),
B.P. 110, F-74941 Annecy--le--Vieux cedex, France\\
\n $^{(b)}$Dipartimento di Fisica dell'Universit\`a di Bologna and INFN
sezione di Bologna, Via Irnerio 46,  40126 Bologna, Italy}

\vskip50mm

\begin{abstract}

We derive and analyze exact static solutions to the gravitating $O(3)$ $\s$ 
model with cosmological constant in (2+1) dimensions. Both signs of 
the gravitational and cosmological constants are considered. Our solutions 
include geodesically complete spacetimes, and two classes of black holes. 

\end{abstract}

\newpage

\section{Introduction}

In a recent paper \cite{CF}, we have constructed and analyzed exact
black hole solutions to the gravitating $O(3)$ $\s$ model in (2+1)
dimensions. Besides Schwarzschild--like black holes, we also found two
numerable families of cold black holes with multiple horizons of infinite
area and vanishing Hawking temperature. It was then natural to extend this
investigation and inquire whether these remarkable cold black hole
solutions survive in the case of a non--vanishing cosmological constant. As
we shall see below, they do not, but are replaced by equally interesting
horizonless, geodesically complete solutions, while the
Schwarzschild--like black holes give way to two distinct families of black
hole solutions. 

Static, rotationally symmetric solutions to the cosmological gravitating 
$\s$ model were previously investigated by Kim and Moon \cite{KM}. They 
integrated numerically the Einstein--$\s$ equations with boundary conditions
appropriate to solitonic configurations, and obtained topological solitons
with integral winding number, as well as non--topological solitons with
half--integral winding number. They also found numerically extreme black
hole solutions as limits of non--topological solitons. The present
investigation is complementary to that of reference \cite{KM}. We shall
consider only solutions following a geodesic in target space. This
approach will enable us to obtain analytically all such geodesic
solutions, without prescribed boundary conditions. These exact geodesic
solutions were actually all excluded from the analysis of Kim and Moon by
their choice of ans\"atze and/or boundary conditions.   

In the next section we introduce the cosmological gravitating $\s$ model,
which is reduced by the geodesic ansatz to cosmological gravity coupled to
a massless scalar field $\s$, or to the dual cosmological
Einstein--Maxwell theory (to which it is equivalent outside sources). In
the case of static rotationally symmetric solutions, the scalar field
$\s$ depends only on one coordinate, either the radial coordinate $\rho$ or
the angular coordinate $\theta$. All the solutions such that $\s =
\s(\rho)$ are derived and analyzed in Sect.\ 3, for both signs of the
cosmological constant and of the gravitational constant. Depending on the
value of the integration constant, these solutions either have naked
singularities, which however are, for a certain parameter range, at
infinite affine distance on timelike or spacelike geodesics. Or they are
geodesically complete, with the wormhole spatial topology; these regular
solutions qualify as non--topological solitons. Sect.\ 4 is devoted to the
discussion of the solutions with $\s = \s(\theta)$. These fall into two
classes. The first class contains the charged BTZ black holes \cite{BTZ}
and related solutions of the cosmological Einstein-Maxwell theory, while
the black holes of the second class are of the form $AdS_2 \times S^1$,
and may be obtained as the near--horizon limits of extreme black holes of
the first class. 

\section{Model equations}

The three--dimensional $O(3)$ non--linear $\s$ model coupled to
cosmological gravity is defined by the action
\be
S=\frac{1}{2}\int d^3 x
\sqrt{|g|}\left[-\frac{1}{\k}(g^{\mu\nu}R_{\mu\nu} + 2\L) +
g^{\mu\nu}\partial_{\mu}{\bm\phi}\partial_{\nu}{\bm\phi}\right] \>,
\label{act}
\ee
where the isovector field ${\bm\phi}$ is valued on the two-sphere
\be
{\bm\phi}^2=\nu^2 \>.
\label{constr}
\ee
We will allow for both signs of the
three--dimensional Einstein constant $\kappa$ (this sign is not
fixed a priori in three--dimensional gravity \cite{DJH} ) and of the
cosmological constant $\L$. The
Euler equations for the $\s$--model field may be written as
\be
D_{\mu}D^{\mu} {\bm\phi} - \lambda {\bm\phi} = 0\ , \label{sigeq1}
\ee
with $D_{\mu}$ the spacetime covariant derivative, and $\lambda$ a Lagrange
multiplier to be determined from the constraint (\ref{constr}).

In this paper we will consider only solutions depending on a single scalar
potential $\s$. Eq. (\ref{sigeq1}) then reduces to
\be
D_{\mu}\s D^{\mu}\s\frac{d^2{\bm\phi}}{d\s^2} - \lambda {\bm\phi}
=
- D_{\mu}D^{\mu}\s\frac{d{\bm\phi}}{d\s}\ .
\label{sigeq2}
\ee
Without loss of generality \cite{NK}, the potential $\s$ may be chosen to
be harmonic,
\be
D_{\mu}D^{\mu}\s = 0 \ , \label{harm}
\ee
so that the $\s$--model field follows a geodesic on the sphere
${\bm\phi}^2=\nu^2$, i.e. a large circle parametrized by the angle $\s$.
The action (\ref{act}) then reduces to that of a massless scalar field
coupled to cosmological gravity
\be\lb{acts}
S=\frac{1}{2}\int d^3 x
\sqrt{|g|}\left[-\frac{1}{\k}(g^{\mu\nu}R_{\mu\nu} + 2\L) +
\nu^2 g^{\mu\nu}\partial_{\mu}\s\partial_{\nu}\s\right]\ .
\ee
As discussed in \cite{CF}, this effective Einstein--scalar theory is
locally equivalent to sourceless Einstein--Maxwell theory in (2+1)
dimensions with a cosmological constant,
\be\lb{actM}
S=\frac{1}{2}\int d^3 x
\sqrt{|g|}\left[-\frac{1}{\k}(g^{\mu\nu}R_{\mu\nu} + 2\L) -
\frac{1}{2}\,g^{\mu\nu}g^{\rho\s}F_{\mu\rho}F_{\nu\s}\right]\>,
\ee
the second group of
Maxwell equations $D_{\nu}F^{\mu\nu} = 0$ being the integrability
condition for the duality relation
\be
\lb{dual} F^{\mu\nu}=
\frac{\nu}{\sqrt{|g|}}\epsilon^{\mu\nu\lambda}\partial_{\lambda}\s\>,
\ee
while the first group of Maxwell equations leads to the harmonicity
condition (\ref{harm}).

Now we specialize to static rotationally symmetric solutions.
A convenient parametrization for the spacetime metric is then
\cite{EL, EML}
\be\lb{el}
ds^2 = U\,dt^2 + V\,d\theta^2 +
\zeta^{-2}\,\frac{d\rho^2}{UV}\>,
\ee
where $\theta$ is an angle
and the metric fields $U > 0$, $V < 0$ and $\zeta$ depend only on
the radial coordinate $\rho$. The correspondence with the
parametrization of Kim and Moon  \cite{KM},
\be
ds^2 = \e^{2N}\,B\,dt^2 - \frac{dr^2}{B} - r^2\,d\theta^2\>,
\ee
is
\be
r^2 = -V\,, \quad \e^{-2N} = \frac{\zeta^2}{4}\dot{V}^2\,, \quad B
= \frac{\zeta^2}{4}\,U\dot{V}^2\,,
\ee
where $\dot{} \equiv
d/d\rho$. In (\ref{el}), the scaling function $\zeta(\rho)$ may and
will be taken equal to 1 after variation. The constraint
$g_{\rho\theta} = 0$ is consistent with the Euler equations for
the action (\ref{acts}) iff
\be
T_{\rho\theta} \equiv \nu^2\partial_{\rho}\s\,\partial_{\theta}\s =0\,,
\ee
leading to the two possibilities $\s = \s(\rho)$, or $\s = \s(\theta)$.

The first possibility $\s = \s(\rho)$ corresponds, in the parametrization of \cite{KM},
to $F(r) = \s(\rho)$, $n = 0$. This case was discussed very briefly (in the more
general stationary case) at the end of \cite{EL}. The variational problem (\ref{acts})
reduces to the one--dimensional problem for the effective Lagrangian
\be
L = \frac{\zeta}{2}\,\left(-\frac{1}{2\kappa}\,\dot{U}\dot{V} +
\nu^2 UV\dot{\s}^2\right) - \frac{\Lambda}{\kappa\zeta}\,.
\ee
The elimination of the cyclic variable $\s$ in terms
of its constant conjugate momentum $p$ by
\be
\dot{\s} = \frac{p}{\nu^2 \zeta UV}
\ee
then leads to the reduced Lagrangian (or, more precisely, the Routhian
\cite{Goldstein})
\be\lb{lag1}
\hat{L}\, \equiv\, L - p\dot{\s}\, = \, -\frac{\zeta}{4\kappa}\,\dot{U}\dot{V}
- \frac{1}{2\zeta}\left(\frac{p^2}{\nu^2 UV}+\frac{2\Lambda}{\kappa}\right)\,.
\ee
Note that the electromagnetic field dual to $\s(\rho)$,
\be\lb{Frho}
F_{t\theta} = \nu p\,,
\ee
is globally defined only if the spatial
topology is such that the loops $\rho =$ constant are
non--contractible. The solutions of the equations of motion
derived from (\ref{lag1}) are discussed in  Sect. 3.

The second possibility $\s = \s(\theta)$ reduces, after using the
harmonicity condition
(\ref{harm}) and the single--valuedness of the $\s$--model field
$\bm\phi$, to
\be
\s = n\,\theta
\ee
($n$ integer), corresponding in the
parametrization of \cite{KM} to $F(r) = 0$. In this case the
constant of motion is the dual electromagnetic field density
\be
\Pi^{t\rho} \equiv -\zeta^{-1}F^{t\rho} = n\nu\,,
\ee
corresponding to the quantized
electric charge $2\pi n\nu$. Accordingly, the Einstein--Maxwell
action (\ref{actM}) reduces to the reduced Lagrangian
\be
\hat{L}\, \equiv \,L - \Pi^{t\rho}F_{t\rho}\, =
- \frac{\zeta}{4\kappa}\dot{U}\dot{V}
- \frac{1}{\zeta}\left(\frac{\L}{\kappa}
- \frac{n^2\nu^2}{2V}\right)\,. \label{cicu}
\ee
The solutions of the
corresponding equations of motion, first derived in \cite{EML},
will be revisited in Sect. 4.

\setcounter{equation}{0}
\section{Non--topological solitons ($\s = \s(\rho)$)}

The Lagrangian (\ref{lag1}) leads  (after setting the scale
$\zeta(\rho) = 1$) to the equations of motion
\ba & V\ddot{U} =
U\ddot{V} = - \frac{\ds 2\kappa p^2}{\ds \nu^2 UV}  \,, &  \lb{mot1} \\
& \dot{U}\dot{V} - \frac{\ds 2\kappa p^2}{\ds \nu^2 UV}  = 4\Lambda \,.&
\lb{ham1}
\ea
The linear combination of (\ref{mot1}) and
(\ref{ham1}) leads to the equations $ (V\,\dot{U})\dot{} =
(U\,\dot{V})\dot{} = 4\Lambda$, which are integrated by
\ba
V\dot{U} & = & 4\Lambda(\rho - \alpha)\,, \lb{u} \\
U\dot{V} & = & 4\Lambda(\rho - \beta)\,, \lb{v}
\ea
where $\alpha$ and $\beta$
are two integration constants. The sum of these two equations may
be further integrated (the integration constant being fixed by the
Hamiltonian constraint (\ref{ham1})) to
\ba
UV & = & 4\Lambda(\rho
- \alpha)(\rho - \beta) - \frac{\kappa p^2}{2\Lambda\nu^2 } \nonumber \\
& = & 4\L(\rho - \rho_+)(\rho - \rho_-)\,, \lb{uv}
\ea
with
\be
\rho_{\pm} = \frac{\alpha + \beta \pm \delta}{2}\,, \qquad \delta
= \sqrt{(\alpha - \beta)^2 + \kappa p^2/2\L^2\nu^2 }\,.
\ee
Finally
equations (\ref{u}) and (\ref{v}) may be combined with (\ref{uv})
to yield equations for $\dot{U}/U$ and $\dot{V}/V$ which are
readily integrated. The resulting metric depends on the sign of
$\delta^2$: \\

\n 1) $\delta^2 > 0$. The metric and scalar field are given by
\ba
ds^2 & = & A\,|\rho -
\rho_+|^{1/2+a}\,|\rho - \rho_-|^{1/2-a}\,dt^2
- \,\frac{4|\L|}{A}\,|\rho - \rho_+|^{1/2-a}\,|\rho
- \rho_-|^{1/2+a}\, d\theta^2 \nonumber \\
& & + \,\frac{d\rho^2}{4\L(\rho - \rho_+)(\rho - \rho-)}\,,
\qquad \s = \frac{p}{4\L\nu^2\delta}\,
\ln(\frac{|\rho-\rho_+|}{|\rho-\rho_-|})\,,
\label{cicca}
\ea
where $A$ is an integration constant (which we will assume to be positive),
and $a = (\beta-\alpha)/2\delta$, with $a^2 > 1/4$ for $\k < 0$, and
$a^2 < 1/4$ for $\k > 0$.
The metric (\ref{cicca}) is Lorentzian and static, with $\partial_t$ as
timelike Killing vector, in the intervals where
the product $UV$ is negative, i.e. in the ranges
$\rho > \rho_+$ or $\rho < \rho_-$ for $\L < 0$, and $\rho_- < \rho < \rho_+$
for $\L > 0$.

In order to study the spacetime
structure we first compute the scalar curvature
\be
R=-6\L + \frac{\k p^2}{4\L\nu^2(\rho-\rho_+)(\rho-\rho_-)}\, \label{scaca}
\ee
which diverges at $\rho=\rho_{\pm}$ and goes to
a constant as $\rho\to\infty$.
The nature of the singularities  $\rho=\rho_{\pm}$ can be understood by
considering the behavior
of the geodesics in their vicinity. The general expression for the
proper time along radial (i.e. $\theta$ = const.) timelike curves is
\be
\tau = \int d\rho\,\frac{\sqrt{-g_{tt}g_{\rho\rho}}}{\sqrt{E^2 - g_{tt}}}
\label{eqgeo}\ .
\ee
Application of the metric (\ref{cicca}) tells us that for instance
the singularity $\rho = \rho_+$ is reached in a finite proper time
only if $a \geq -1/2$, whereas for $a<-1/2$ (which is possible only
for $\k < 0$) there is always a
turning point at some $\rho_1>\rho_+$, and $\rho_+$ is never
reached. For null geodesics the affine parameter
\be
v=\frac{1}{E}\int d\rho \sqrt{-g_{tt}g_{\rho\rho}} \label{geonu}
\ee
goes to infinity for $\rho \to \rho_+$ provided that $a<-3/2$,
and is bounded otherwise. Similar considerations apply for the singularity
at $\rho = \rho_-$ (note that the solution (\ref{cicca}) is invariant
under the involution $\rho_+ \leftrightarrow \rho_-\,, a \leftrightarrow -a\,,
p \leftrightarrow -p$).
In the case $\L < 0$, the other boundary of the spacetime, $\rho = \pm\infty$,
is null and spacelike complete, while all timelike geodesics bounce inward at
some finite $\rho_2$. Therefore, despite the behavior of $R$ in
eq. (\ref{scaca}), for $a<-3/2$ (resp. $a>3/2$) the whole spacetime $\rho > \rho_+$
(resp. $\rho < \rho_-$) is null and
timelike complete and so physically regular (only spacelike
trajectories 'feel' the singularity at the point $\rho=\rho_\pm$).
The global behavior of
timelike geodesics, always bounded between $\rho_1$ and $\rho_2$,
and the existence of two timelike regular boundaries
($\rho=\rho_{\pm}$ and $\rho=\pm\infty$) are exactly what happens also in
pure AdS spaces. As in those cases, the full Penrose diagram is an
infinite strip extending in the vertical $t$ direction.

We find it interesting to comment on some limiting
cases. The limits  $a \to \pm 1/2$, or $\beta-\alpha \to \pm\delta$,
correspond to $\k p^2 \to 0$, i.e. the $\s$-model  field decouples.
Then  $R=-6\L$  and all curvature singularities disappear.
In particular, $a=1/2$ is nothing but the
BTZ black hole \cite{BTZ} ($\rho_+$ is the location of the event
horizon) and $a=-1/2$ the Deser-Jackiw-\linebreak 't Hooft solution
\cite{DJT}. The connection with the corresponding solutions of the
gravitating $\s$--model in the case $\L=0$ (see \cite{CF})
is more involved. The limit $\L\to 0$ can be performed only locally, e.g.
in the vicinity of $\rho_+$. Starting from the metric
(\ref{cicca}) let us introduce the coordinates $(x,\bar t)$ via
($c$ is an appropriate constant)
\be
\rho=\rho_+ +  c|\L| x^{4/(1+2a)},\ \ \ \bar t = |\L| t\ .
\ee
and take the limit $\L\to 0$ while keeping the
quantity $A|\L|^{1/2+a}=$ const (the relation between $a$ and
the parameter $\alpha$ used in  \cite{CF} is $\alpha=(1-2a)/(1+2a)$).
In particular,
for $a<-3/2$ and in the limit $x\to 0$ a horizon appears and,
therefore, a black hole by analytic continuation to negative
values of $x$. These features were absent in the original metric
(\ref{cicca}).\\

\n 2) $\delta^2=0$. In this case (which can occur only for $\k < 0$),
the two singularities coincide,
$\rho_+=\rho_-\equiv \rho_0$, so that the spacetime metric can be Lorentzian
and static only for $\L < 0$. Invariance under translations in $\rho$
allows us to take $\rho_0 = 0$, leading to the form of the solution
\ba
ds^2 & = & A |\rho|e^{b/2\rho}\,dt^2
+ \frac{4\L}{A}\,|\rho|e^{-b/2\rho}
\,d\theta^2 + \frac{d\rho^2}{4\L\rho^2}\,, \nonumber \\
\s & = & -\frac{p}{4\L\nu^2\rho}\,, \label{ridu}
\ea
with $A > 0$ and $b = (-\k p^2/2\L^2\nu^2)^{1/2}$ (we have taken into account
the involution mentioned above to keep only the positive root),
and the scalar curvature
\be
R = - 6\L + \frac{\L b^2}{2\rho^2}\ .\label{scadu}
\ee
The nature of the singularity at $\rho = 0$ depends on the range ($\rho > 0$
or $\rho < 0$) considered. For $\rho > 0$, it
is not only null, but also spacelike complete,
whereas timelike geodesics always bounce back and never reach it.
The global causal structure
of this regular spacetime is the same as for the solutions $a < -3/2$
analyzed in case 1 (the present case corresponds to $a \to -\infty$). For
$\rho < 0$ all geodesics terminate at $\rho = 0$ for a finite affine
parameter, i.e. the spacetime is truly singular there. Finally, for
$b = 0$ the geometry becomes regular and corresponds to the BTZ vacuum (the
quantity $\delta^2$ measures the mass of the solution). \\

\n 3) $\delta^2<0$. Again in this case necessarily $\k < 0$ and $\L < 0$.
Putting $\eta^2 \equiv -\delta^2/4$, and choosing $\alpha + \beta = 0$,
we obtain the solution
\ba
ds^2 & = & A \sqrt{\rho^2 + \eta^2}
\,\exp[-\tilde{\alpha}\arctan(\rho/\eta)]\,dt^2
 \nonumber \\
& & + \,\frac{4\L}{A}\sqrt{\rho^2 + \eta^2}
\,\exp[\tilde{\alpha}\arctan(\rho/\eta)]\,d\theta^2\,
+ \,\frac{d\rho^2}{4\L(\rho^2 + \eta^2)}\;, \label{metre} \\
\s & = & \frac{p}{4\L\nu^2\eta}\,\arctan(\rho/\eta) \nonumber
\ea
($\tilde{\alpha} = \alpha/\eta$), and the corresponding scalar curvature
\be
R=-6\L +\frac{\k p^2}{4\L\nu^2(\rho^2+\eta^2)} \,. \label{scatre}
\ee
This solution is everywhere regular in the whole range
$-\infty<\rho<+\infty$. It is a non-topological soliton, disproving
the claim of nonexistence of such $n = 0$ regular solutions
made by Kim and Moon \cite{KM}. This claim relied on the occurrence of a
logarithmic divergence of $F(r) \equiv \s(\rho)$ at $r^2 \equiv
-g_{\theta\theta} = 0$; however our solution (\ref{metre}) has the
wormhole topology, with $-g_{\theta\theta}$ positive everywhere. \\

This completes the analysis of the regular solutions of the $\sigma$-model
equations for $\sigma=\sigma(\rho)$. Let us add that the last two cases
($\Lambda < 0$, $\k < 0$ and $\delta^2 \le 0$), which lead to geodesically
complete spacetimes with the topology $R^2 \times S^1$, also correspond to
fully regular electrostatic solutions of the cosmological 
Einstein--Maxwell theory (\ref{actM}), with closed lines of force from
(\ref{Frho}).  

\setcounter{equation}{0}
\section{The black hole class ($\s = \s(\theta)$)}

We now turn to the analysis of the Lagrangian (\ref{cicu}). The equations
of motion, after setting $\zeta(\rho)=1$, are
\ba
\ddot{V} & = & 0 \,, \lb{eupo} \\
\ddot{U} & = & {\ds\frac{2\k n^2\nu^2}{V^2}}  \,, \lb{eddo} \\
\dot U\dot V & = & 4\L -{\ds\frac{2\k n^2\nu^2}{V}} \,. \lb{edco}
\ea
Eq. (\ref{eupo}) is integrated by $V = a\rho + b$ ($a$, $b$ constants),
leading to two cases (see also \cite{EML}).\\

1) In the generic case $a \neq 0$, we can always translate the radial
coordinate $\rho$ and rescale the time coordinate $t$ so that $V=-2\rho$
($\rho \ge 0$). Then the metric is given in Schwarzschild form by
\be
\label{uppo}
 ds^2 = -\L (r^2- 2\gamma^2\ln(r/r_0))\,dt^2 -
r^2\,d\theta^2 + \frac{dr^2}{\L(r^2- 2\gamma^2\ln(r/r_0))}\,,
\label{care} \ee with $r^2 = 2\rho$, and $\gamma^2 = -\k
n^2\nu^2/2\L$. The Ricci scalar is
\be
R=-2\L\left(3 -\frac{\gamma^2}{r^2}\right)\,,
\ee
indicating the presence of a curvature singularity at
$r=0$.

For $\L < 0$, we recognize in (\ref{care}) the electrically charged BTZ
solution \cite{BTZ}. The horizons are defined as the surfaces where
$f(r) \equiv -\L(r^2- 2\gamma^2\ln(r/r_0) = 0$. Let us first
consider the case $\k>0$ ($\gamma^2 > 0$).Then $f(r)$ is minimum for $r =
\gamma$. We may distinguish the following three different cases:

\noindent i) $\gamma > \sqrt{e} r_0$ ($f(\gamma)<0$). This means
that $f(r)$ vanishes for two values $r_{\pm}$ of $r$, $r_+$ being
the black hole horizon and $r_-$ the inner horizon.

\noindent ii) $\gamma = \sqrt{e} r_0$ ($f(\gamma)=0$). This is the
extremal solution where the two horizons merge $r_+ = r_- =
\gamma$.

\noindent iii) $\gamma < \sqrt{e} r_0$ ($f(\gamma)>0$). In this
case there are no horizons and the singularity is naked.

\noindent In these three cases the causal stucture of the solution
(\ref{care}) is the same as in the correponding cases of the
4--dimensional Reissner-Nordstr\"om-Anti-de Sitter (RN-AdS) solution.
Furthermore, it can also be verified that all geodesics behave as in
the RN-AdS spacetime.

On the other hand, in the case $\k<0$ ($\gamma^2 < 0$), the equation
$f(r)=0$ has always one solution, so that there is a single horizon, and
all geodesics terminate at the spacelike singularity $r=0$ in a finite
proper time and affine parameter. This spacetime is the analogue of the
Schwarzschild-Anti-de Sitter (or of the static BTZ) spacetime.

Let us now consider the case of a positive cosmological constant, $\Lambda
> 0$. Then, for $\k > 0$ ($\gamma^2 < 0$), there is always one horizon at
$r = r_h$, with $f(r) > 0$ for $0 < r < r_h$. The Penrose diagram
is obtained by a $\pi/2$ rotation from that 
of the Schwarzschild-Anti-de Sitter black hole. For $\k < 0$
($\gamma^2
> 0$), the spacetime contains sectors where the the Killing vector
$\partial_t$ is timelike only if $\gamma > \sqrt{e} r_0$. These
sectors are connected by two horizons with cosmological sectors
terminating respectively at the spacelike singularity $r = 0$ and
at spacelike infinity. The Penrose diagram is similar to that of the
Reissner-Nordstr\"{o}m-Anti-de Sitter black hole, rotated by $\pi/2$
(or of the Schwarzschild-de Sitter black hole, including the case
$\gamma=\sqrt{e}r_0$ where the black hole and the de Sitter
horizons merge) .
\\

2) The second case  is given by $V$ constant,
leading from Eq. (\ref{edco}) to $V = - \gamma^2$.  Then eq.
(\ref{eddo}) can be solved and the general form of the metric is
\be\label{uppi}
 ds^2 = -\frac{2\L}{\gamma^2}(\rho^2 +c)\,dt^2 -
\gamma^2\,d\theta^2 + \frac{d\rho^2}{2\L(\rho^2 +c)}\ ,
\ee 
where $c$ is an integration constant. This
metric is Lorentzian for $\k\L < 0$ if the range of $\rho$ is such
that $\L(\rho^2+c) < 0$. The mixed Ricci tensor components are
constant, $R^t_t = R^{\rho}_{\rho} = -2\L, R^{\theta}_{\theta} =
0$, showing that the geometry is regular for all $\rho$.

The causal structure depends on the signs of $\L$ and $c$. For $\L
< 0$ ($\k > 0$), the metric (\ref{uppi}) is Lorentzian for
$\rho^2 + c > 0$. If $c>0$, the solution is globally $AdS_2 \times
S^1$. If $c<0$, then the spacetime is the direct product of the
$AdS_2$ black hole \cite{JT} (or extreme black hole if $c = 0$) of
mass $|c|$ with the circle $S^1$. This spacetime has two horizons
at $\rho_{\pm}=\pm\sqrt{|c|}$ and two asymptotic regions
$\rho=\pm\infty $ . Its boundaries are null and spacelike
complete, while timelike geodesics cross both horizons and are
bounded between $\rho_1$ ($<\rho_-$) and $\rho_2$ ($>\rho_+$).
Finally, for $\L > 0$ ($\k < 0$), then the metric (\ref{uppi}),
is Lorentzian for $\rho^2 + c < 0$ (implying $c < 0$), and is the
3--dimensional version of the Nariai solution, i.e.\ the direct 
product of the 2--dimensional de Sitter spacetime $dS_2$
with the circle $S^1$.

The enhanced symmetry of these black hole solutions suggests that 
they may be obtained as near horizon geometries of the generic 
black holes of case 1)  in the limit 
where two horizons become coincident. To show this (a similar
proof for generic 2D dilaton gravity theories is presented in
\cite{CFNN} and for $D\geq 4$ Einstein-Maxwell theory in
\cite{NNN}), let us start with the generic black hole metric 
(\ref{uppo}) and define
\be
r_0 \equiv \gamma {\e}^{-1/2 +c\alpha^2/\gamma^2}\, , \quad
t=\frac{\tilde t}{\alpha\gamma}\,, \quad  
r=\gamma +\alpha\rho \, .
\ee
When the near-horizon limit $\alpha\to 0$ ($r\to\gamma$) is taken 
(for either $\Lambda<0$, $\k>0$ or $\Lambda>0$, $\k<0$), the 
metric (\ref{uppi}) is obtained.

\vspace{0.5cm}

\noindent {\bf Acknowledgement:}
A large part of this work was carried out while G.C. was at the
Laboratoire de Gravitation et Cosmologie Relativistes (LGCR), Paris,
France, and A.F. at the Physics Department of Stanford University, 
USA.

\end{document}